\documentclass[manuscript]{aastex6}
\usepackage{graphicx}
\usepackage{subfigure}
\usepackage{longtable}
\usepackage{multirow}

\shorttitle{}
\shortauthors{Xu \& Li}

\begin{document}

\title{A \textit{Chandra} Study of The Stellar X-ray Emissivity of Globular Clusters in M\,31 Bulge}

\author{Xiao-jie Xu}
\affil{School of Astronomy and Space Science, Key Laboratory of Modern Astronomy and Astrophysics, Nanjing University, Nanjing, P. R. China 210046}
\email{xuxj@nju.edu.cn}
\author{Zhiyuan Li}
\email{lizy@nju.edu.cn}
\affil{School of Astronomy and Space Science, Key Laboratory of Modern Astronomy and Astrophysics, Nanjing University, Nanjing, P. R. China 210046}

\begin{abstract}
The X-ray emissivity (i.e., luminosity per unit stellar mass) of globular clusters are an important indicator of their dynamical evolution history. 
Based on deep archival \textit{Chandra} observations, we report a stacking analysis of 44 globular clusters (GCs) with 0.5-8 keV luminosities $L_{\rm X} \lesssim 10^{35} {\rm~erg~s^{-1}}$ in the M\,31 bulge, which are supposed to be dominated by cataclysmic variables (CVs) and coronally active binaries (ABs). 
We obtain a significant detection at $5\sigma$ level in 0.5-8 keV band. The average X-ray luminosity per GC and the average X-ray emissivity are determined to be $5.3 \pm 1.6\times10^{33}\rm~erg~s^{-1}$ and $13.2\pm4.3\times10^{27}\rm~erg~{\rm s^{-1}~M^{-1}_{\odot}}$, respectively. Both of these values are consistent with those of MW GCs. What's more, the measured emissivity of M31 GCs is also consistent with that of the MW field stars. Massive GCs have X-ray luminosities which are marginally higher with less massive ones. Massive GCs also show a lower emissivity ($4.5\pm 2.4\times10^{27}{\rm~erg~s^{-1}~M^{-1}_{\odot}}$) than less massive ones ($15.0\pm 7.8\times10^{27}{\rm~erg~s^{-1}~M^{-1}_{\odot}}$), which is consistent with the scenario that the (progenitors of) CVs and ABs were more efficiently destructed via stellar encounters in the more massive GCs. No dependence of the X-ray emissivity on GC color or on the projected galactocentric distance of GCs were found.
\end{abstract}

\section{Introduction}
Globular clusters (GCs) in our Galaxy are rich in stellar X-ray sources. The identified sources include (quiescent) low-mass X-ray binaries ((q)LMXBs), millisecond pulsars (MSPs), cataclysmic variables (CVs) and coronally active binaries (ABs). These sources are mostly located inside the half-light radius ($r_{\rm h}$) of GCs \citep[e.g.,][]{dis02,poo03,kim06,zha11,dag14, czq17}. Compared to the Galactic field, GCs in both the Milky Way (MW) and external galaxies were found to have much higher chances to host LMXBs with a luminosity $L_{\rm X} \gtrsim 10^{35}$, which are thought to be formed via close stellar encounters in the crowded environment in GCs \citep{fab75,pfa02,poo03,iva05,iva08,hur07,fre09,bah13,aga13}. The majority of GCs, on the other hand, have $L_{\rm X}$ below $10^{34-35} {\rm~erg~s^{-1}}$, which are supposed to be dominated by CVs and ABs \citep[e.g.,][]{poo03,czq17}. Compared to LMXBs, the progenitors of CVs and ABs are less massive and have a longer evolution timescale.
Hence, dynamical processes are also expected to affect the formation of CVs and ABs in GCs. 
Indeed, previous work \citep[e.g.,][]{poo03,poo06,max12} revealed a correlation between the number of weak X-ray sources detected in GCs and the so-called stellar encounter rate, which was interpreted as evidence for a dynamical origin for such sources, in particular CVs. 
However, stellar interactions, 
including two-body and three-body encounters, should take place with competing effects, in which binaries can be created in two-body interactions, but also can be destroyed or modified in three-body interactions (Hut et al. 1992).
In this regard, whether the number of CVs and ABs could be effectively elevated by the stellar encounters in GCs remains an open issue. 

One way of testing the net outcome of the dynamical interactions is to compare the X-ray emissivities (luminosity per unit stellar mass, $\varepsilon_{\rm X}$) of GCs with that of field stars. Recently, we carried out such a study on 69 MW GCs, obtaining an average $\varepsilon_{\rm X}$ of $\sim 7.3 \pm 2.7 \times 10^{27}{\rm~erg~s^{-1}~M_{\odot}^{-1}}$ in the 0.5-8 keV band \citep{czq17}. This is found to be lower than the field level, which is represented by CVs and ABs detected in the Solar neighborhood and the cumulative X-ray emission from gas-poor dwarf elliptical galaxies in the Local Group ($\sim 12\times10^{27} {\rm~erg~s^{-1}~M_{\odot}^{-1}}$; \citealt{saz06,gec15,czq17}). This provides strong evidence for dynamical destruction of the progenitors of CVs and ABs, due chiefly to binary-single interactions in GCs \citep{czq17}.

It is desired to extend the above study to GCs in other galaxies. However,
unlike the luminous LMXBs (with $L_{\rm X} \gtrsim 10^{36}{\rm~erg~s^{-1}}$) that have been routinely detected in external galaxies (Fabbiano 2006), the weak X-ray populations studied in \citet{czq17}, when placed at extragalactic distances, are beyond the detection sensitivity of current X-ray facilities.
Moreover, even a single LMXB with $L_{\rm X} \approx 10^{35}{\rm~erg~s^{-1}}$ can easily mask the cumulative X-ray emission from the numerous CVs and ABs in the same host GC.
In this regard, Local Group galaxies are perhaps the only laboratory to study the properties of weak X-ray sources in GCs. 
In particular, M\,31, the nearest massive galaxy with a large GC population, is the best-suited target. M\,31 has been observed by \textit{Chandra} for more than 100 times with an accumulated exposure approaching 1 Ms. This ensures a source detection limit down to a few times $10^{34} {\rm~erg~s^{-1}}$ in its bulge, thus allowing for a clean identification of GCs hosting bright LMXBs \citep{kon02,fan05,pea10b,bar14} and opening up the opportunity of measuring the cumulative emission from the fainter stellar populations, i.e., CVs and ABs. 

In this work, we present a study on confirmed GCs in the M\,31 bulge with individual $L_{\rm X}$ below the \textit{Chandra} detection limit, by performing a stacking analysis.
The stacking technique has been proved powerful to detect the cumulative X-ray emission of faint, unresovled sources to enable the study of their average properties, e.g., X-ray luminosities of high-redshift galaxies \citep[e.g.,][]{zin12,bas13}. 
Recently, \citet{vul14}, in a study primarily focusing on the resolved X-ray binaries in the bulge and disk of M\,31, performed a stacking analysis on 54 GCs, which was reported to be a non-detection at an upper limit of $\sim10^{32} {\rm~erg~s^{-1}~GC^{-1}}$. 
However, this is puzzling result, in regard with the typical luminosity of a few $10^{33} {\rm~erg~s^{-1}}$ find in the MW GCs (e.g., \citealt{czq17}), 
and is to be contrasted with our present study.

The remainder of this paper is organized as follows. We introduce the data and analysis method in \S~2 and present the results in \S~3.  We discuss the implications of the results and conclude in \S~4. Throughout this work we adopt a distance of 780 kpc for the M\,31 GCs, and a Galactic foreground absorption column density 0f $N_{\rm H}=7.0\times10^{20}{\rm~cm^{-2}}$. 
Quoted errors are at 90\% confidence level, unless otherwise stated.

\section{Data \& Analysis}

\subsection{X-ray Data Preparation}

\begin{figure}[htbp]
\centering
\includegraphics[height=4in,width=5.4in]{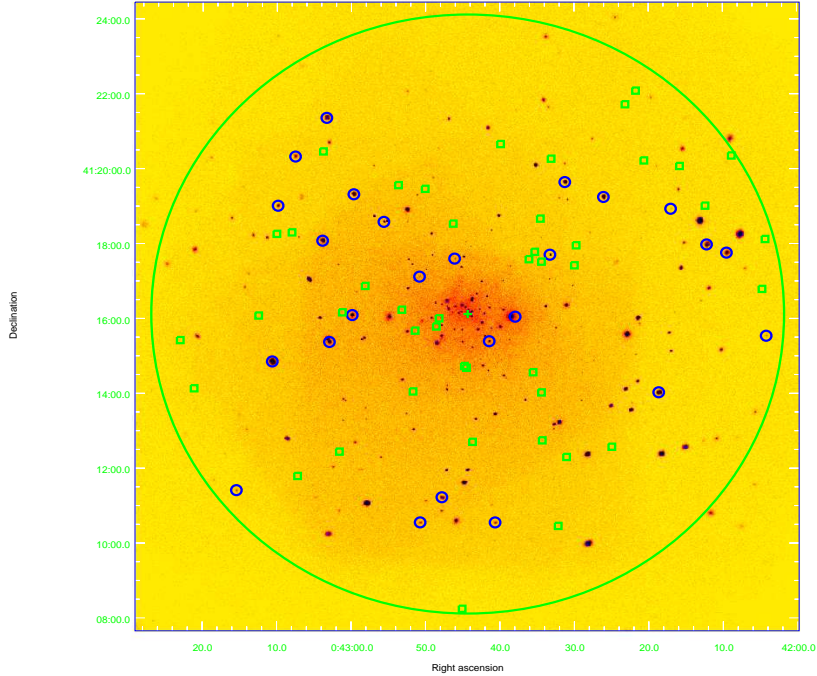}
   \caption{0.5-8 keV merged \textit{Chandra} ACIS image of the M\,31 bulge, overlaid by 25 X-ray-detected (blue circles) and 44 non-detected (green squares) GCs. The latter are included in stacking analysis. The large circle measures a galactocentric radius of 8$^\prime$, while the cross marks the M\,31 center.}
   \label{fig:01}
\end{figure}

The central $\sim8^\prime$ region of M\,31 is one of the most frequently visited targets of \textit{Chandra}. For our purpose, we utilize 122 publicly available ACIS observations taken between 1999 to 2013 (including 81 ACIS-I and 12 ACIS-S observations), which were primarily for monitoring the X-ray binary populations in the bulge (e.g., Barnard et al.~2014). 
Individual level 1 data files from each observation were reprocessed to produce the level 2 data files using CIAO v.4.5 and the corresponding calibration files, following the standard pipeline. 
For the ACIS-S observations, we used only data taken from the S2 and S3 CCDs.
Counts, exposure and point-spread function (PSF, defined as $90\%$ encircled energy radius) maps in the 0.5-2 (S), 2-8 (H), and 0.5-8 (F) keV bands were created. 
A first run of the CIAO tool {\it wavdetect} was employed to detect and locate sources in each observation. We then corrected for the relative astrometry of each observation with respect to ObsID 1575, which has the longest exposure, by matching centroids of the commonly found bright sources. The counts and exposure maps were reprojected to produce the final, merged images. The PSF maps were also merged, individually weighted by the exposure map at a given pixel. 
The total effective exposure exceeds 700 ks in the central $\sim2^\prime$ and gradually decreases to a value of $\sim 400$ks at radii out to 8\arcmin.
The merged \textit{Chandra} 0.5-8 keV image is presented in Figure \ref{fig:01}.

{\it wavdetect} was employed a second time on the merged images to detecte sources in all three bands. Our final source list consists of 406 independent sources inside 8\arcmin. Further studies of these sources will be presented elsewhere (see aslo Vulic et al.~2016 for an updated {\it Chandra} catalog of X-ray sources in M\,31).    
We inspected the sensitivity map produced in {\it wavdetect} and found that the local detection limit reaches $6.0 \times 10^{-5} {\rm~cts~ s}^{-1}$ within a galactocentric radius of 2\arcmin~ and gradually decreases to $2.0 \times 10^{-4} {\rm~cts~s}^{-1}$ at 8\arcmin. Assuming a fiducial absorbed power-law spectrum with a photon-index of 1.7, we obtain a count rate-to-intrinsic luminosity conversion factor of $9.1 \times 10^{38}{\rm~erg~s^{-1}}/({\rm cts~s^{-1}})$ in the 0.5-8 keV band, and the above limits translate to 0.55$\times10^{35}{\rm~erg~s^{-1}}$ and 1.8$\times10^{35}{\rm~erg~s^{-1}}$, respectively. 
The latter is set as the global detection limit for subsequent analysis. 
(There are two GCs which were below the detection limit, but were detected in the central region since the detection limit is lower in that region. However, they were both excluded from the stacking list because there were contaminations near them.)

\subsection{GC Sample Selection}
The stacking analysis requires a sample of GCs with $L_{\rm X}$ below the detection limit of the merged \textit{Chandra} observations, since any detected X-ray counterpart is expected to be dominated by LMXBs. We start from the Revised Bologna Catalogue of M\,31 globular clusters V.5 by \citet[] [http://www.bo.astro.it/M31/] {gal04} and the GC list by \citet{pea10a}. As the first step, only confirmed GCs within 8\arcmin~ from the M\,31 center are included to make a balance between the number of GCs and the increasing PSF size. This leads to a preliminary sample of 81 GCs. 
Next, this sample is cross-correlated with our X-ray source list. An X-ray source is considered to be associated with a GC if their projected offset is within 1\arcsec~ or two times the source positional uncertainty, whichever is larger. A total of 25 GCs are thus found with an X-ray counterpart (marked as circles in Figure \ref{fig:01}). 
The mean offset of 22 out of the 25 pairs is found to be $0.19\pm0.10\arcsec$.
The other three pairs, located at far off-axis, show large offsets of $\sim5$\arcsec, which could be chance alignments. 
These 25 GCs, along with an addition of 11 GCs located within 3 times the 90\% PSF of a nearby X-ray source, are excluded from further analysis. 
As the last step, we visually examine the vicinity of the remaining 45 GCs and further remove 1 GC that could still be contaminated by nearby bright X-ray sources.  
Our final sample of 44 GCs are presented in Table 1 and marked by squares in Figure \ref{fig:01}.

\begin{table}[h]
\caption{Basic properties of sampled GCs.}
\label{tbl-01}
\tiny
  \begin{center}\begin{tabular}{cccccccc}
  \hline\hline

 GC Name & R.A. & Dec. & $r_{\rm h}$ & $m_{\rm v}$ & $\log~M$ & $m_{\rm b}-m_{\rm v}$ & D\\
       &      &        & (\arcsec)   & &  ($M_\odot$) &  & (\arcmin)\\
\hline              
B075 & 10.536767 & 41.339236 & - & 17.33 & 5.51 & 0.92 & 7.89	 \\ 
B080 & 10.551537 & 41.316914 & - & 17.442 & 5.46 & 1.221 & 6.65	 \\ 
B091 & 10.590458 & 41.368167 & - & 17.56 & 5.14 & 0.80 & 7.31	 \\ 
B093 & 10.596392 & 41.362103 & - & 16.87 & 5.64 & 0.97 & 6.86	 \\ 
B103 & 10.623879 & 41.299339 & - & 15.23 & 6.3 & 1.02 & 3.29	 \\ 
B104 & 10.624883 & 41.290433 & - & 17.51 & 5.21 & 0.77 & 2.99	 \\ 
B106 & 10.629317 & 41.205108 & - & 16.03 & 6.03 & 0.96 & 4.58	 \\ 
B109 & 10.634029 & 41.174408 & 0.8 & 16.22 & 5.58 & 1.12 & 6.12	 \\ 
B114 & 10.642929 & 41.212514 & 0.89 & 17.28 & 5.38 & 0.42 & 3.88	 \\ 
B115 & 10.64335 & 41.233853 & 0.46 & 16.03 & 6.25 & 1.02 & 2.82	 \\ 
B119 & 10.650337 & 41.293164 & - & 17.408 & 5.28 & 0.94 & 2.12	 \\ 
B126 & 10.682004 & 41.211861 & - & 17.149 & 5.32 & 0.77 & 3.43	 \\ 
B127 & 10.685408 & 41.244836 & 1.45 & 14.467 & 6.45 & 0.49 & 1.45	 \\ 
B132 & 10.714196 & 41.26135 & 0.27 & 17.739 & 5.07 & 0.92 & 1.41	 \\ 
B134 & 10.715325 & 41.234283 & 0.41 & 16.57 & 5.6 & 0.91 & 2.50	 \\ 
B136 & 10.723529 & 41.326144 & - & 17.005 & 5.64 & 0.97 & 3.85	 \\ 
B145 & 10.756588 & 41.207469 & 0.29 & 18.1 & 5.12 & 0.32 & 4.92	 \\ 
B152 & 10.791733 & 41.304386 & - & 16.16 & 5.82 & 0.91 & 5.27	 \\ 
B154 & 10.801896 & 41.268022 & - & 16.758 & 5.9 & 0.95 & 5.29	 \\ 
B167 & 10.838017 & 41.235592 & 0.46 & 17.41 & 5.29 & 1.02 & 7.20	 \\ 
B169 & 10.845838 & 41.257042 & - & 17.08 & 6.24 & 1.23 & 7.30	 \\ 
B262 & 10.708542 & 41.324478 & - & 17.605 & 5.2 & 0.79 & 3.50	 \\ 
B264 & 10.721621 & 41.270667 & 0.32 & 17.577 & 5.13 & 1.00 & 1.67	 \\ 
B268 & 10.779967 & 41.196614 & 0.37 & 18.314 & 4.95 & 1.00 & 6.11	 \\ 
NB16 & 10.637892 & 41.337892 & - & 17.55 & 5.23 & 0.66 & 4.64	 \\ 
NB17 & 10.643329 & 41.292061 & - & 18.922 & 4.60 & 0.71 & 2.32	 \\ 
B053D & 10.603958 & 41.209639 & - & 19.761 & 4.26 & 0.88 & 5.10	 \\ 
NB29 & 10.647079 & 41.296439 & - & 18.821 & 4.64 & 1.23 & 2.36	 \\ 
NB35 & 10.643963 & 41.311214 & - & 19.627 & 4.32 & 0.92 & 3.13	 \\ 
NB39 & 10.702287 & 41.263217 & 0.75 & 17.941 & 4.99 & 0.28 & 0.87	 \\ 
NB41 & 10.70075 & 41.266833 & - & 18.097 & 4.93 & 0.35 & 0.74	 \\ 
NB89 & 10.686579 & 41.245614 & - & 17.965 & 4.98 & 0.12 & 1.41	 \\ 
AU010 & 10.742179 & 41.281242 & - & 17.506 & 5.17 & 0.98 & 2.69	 \\ 
B040D & 10.517892 & 41.302036 & - & 18.811 & 4.64 & 0.69 & 7.78	 \\ 
B041D & 10.519667 & 41.279817 & - & 18.319 & 4.97 & 1.26 & 7.47	 \\ 
B064D & 10.64805 & 41.242839 & - & 16.47 & 5.94 & 0.74 & 2.28	 \\ 
B068D & 10.66625 & 41.344422 & - & 18.652 & 4.71 & 0.90 & 4.60	 \\ 
B090D & 10.754983 & 41.269497 & - & 17.373 & 5.47 & 0.98 & 3.17	 \\ 
BH23 & 10.765679 & 41.341106 & 0.32 & 18.828 & 4.64 & 0.74 & 5.66	 \\ 
B523 & 10.692829 & 41.308997 & - & 18.677 & 4.70 & 0.90 & 2.43	 \\ 
DK054A & 10.687808 & 41.137528 & - & 18.439 & 4.79 & 0.90 & 7.89	 \\ 
PHF6-1 & 10.78325 & 41.305061 & - & 17.643 & 5.49 & 0.98 & 4.94	 \\ 
B539 & 10.565833 & 41.334597 & - & 19.37 & 4.42 & - & 6.65	 \\ 
B540 & 10.585833 & 41.337089 & - & 18.89 & 4.61 & - & 6.04	 \\
\hline
\end{tabular}
\end{center}
Note.-Column 1:GC Name. Column 2-3: R.A. and Dec. of GCs in degrees. Column 4: half light radius in arc seconds; Column 5: V-band magnitude. Column 6: $\log M$ in solar units from \citet{maj15}; Column 7: color of GCs (represented by $=m_{\rm b}-m_{\rm v}$, corrected for extinction); Column 8: distance to M\,31 center in arc minutes.
\end{table}

\subsection{Stacking Analysis} \label{subsec:stack}
The stacking analysis is restricted to a 20-pixel$\times$20-pixel ($\sim~10\arcsec\times10\arcsec$) square region centering at the individual sample GCs. 
It is straightforward to stack the counts map (C map) and exposure map in the F, S and H bands. 
The effective exposure in the stacked map is $E = 19$ Ms, with negligible variation within the small area. 
In addition, we make stacked maps for a control field, namely shifting the 44 boxes by a distance of 20 pixels in a source-free, otherwise random direction.

To determine the average net count rate, we consider the central 6-pixel$\times$6-pixel box of the C map as the source region and the rest of the C map as the background region. 
The size of the source region, which obviously affects the net count rate, is chosen for the following considerations. Ideally, the source region should be large enough to enclose stars located with the half-light radius ($r_{\rm h}$). Unfortunately, most of our sample GCs do not have $r_{\rm h}$ available in the literature (see Table 1). Therefore we take the well determined $r_{\rm h}$ of MW GCs (Harrison 2010), and calculate their apparent half-light radii when placed at the distance of M\,31. 
In this way, the characteristic $r_{\rm h}$ of the M\,31 GCs is estimated to be $0\farcs76 \pm 0\farcs42$, or $1-3$ ACIS pixels.
Considering the relative astrometric uncertainty in the X-ray source/GC catalogs, as well as the increased PSF size at far off-axis, a 6-pixel$\times$6-pixel box is a reasonable choice. Moreover, this choice is also supported by the the profile of stacked count map in \S~3.

\section{Results}
The counts inside the defined source region are summed to give the count numbers of the source, while the background is determined by the averaged counts in the background region (blue line in Figure 2, see following description for details on the background determination). Then the net count number is calculated by:

$C_{\rm N}=C_{\rm S}-C_{\rm B}\times A_{\rm S}/A_{\rm B}$ ,

where $C_{\rm N}$, $C_{\rm S}$, $C_{\rm B}$, $A_{\rm S}$ and $A_{\rm B}$ are net counts, counts of the source and the background regions, area of the source and background regions, respectively.

\begin{figure}
\centering
\includegraphics[height=5in,width=4in]{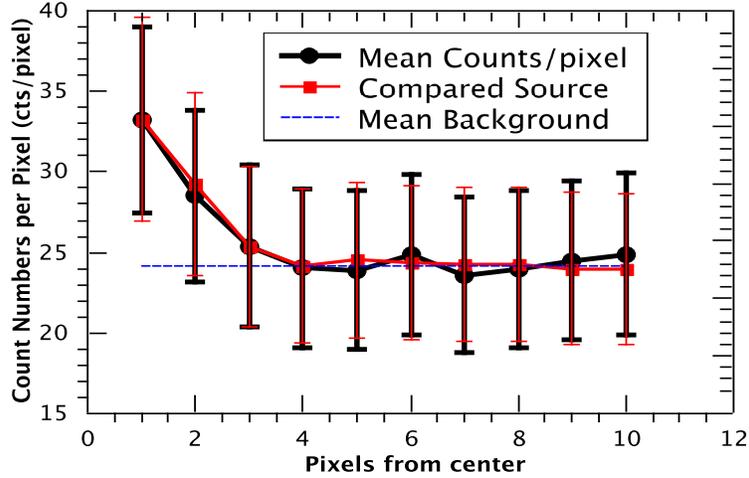}
   \caption{Stacked count profile of the sample GCs in the 0.5-8 keV band, shown by the black dots and black solid line. The blue horizontal line marks the background level determined by averaging the count numbers outside the central $6\times6$ pixels. The red squares and red line are the counts profile obtained by stacking 10 detected sources, normalized to have the same counts in the center pixel as the GC profile.}
   \label{Fig:02}
\end{figure}

In Figure \ref{Fig:02} we present the stacked net C maps of all 44 GCs in F, S, H bands and the control group in F band. The images were smoothed for plotting only. There are 1023 counts (F band) in the source region of the C map, and the respective total background counts in the source region is 871 counts. This revealed a source with $(1023-871)/\sqrt{871}\sim 5 \sigma$ significance in F band. Similar detections occur in S ($\sim 4 \sigma$) and H($\sim 2 \sigma$) bands. The average count rates and luminosities of sampled GCs are given in Table \ref{tbl-02}. The errors of the values is a combination of Poisson error of the source count number, the background count number, and a 10\% additional error to account for the uncertainties brought by background determination and count rate to flux conversion. Comparing to previous detection limit of $\sim~1.4 \times 10^{35} {\rm ergs}~{\rm s}^{-1} $(F band, the same hereafter), the stacking results in an averaged luminosity of $5.3 \pm 1.6\times10^{33}\rm {ergs}~{\rm s}^{-1} {\rm GC}^{-1}$. This is more than an order of magnitude improvement. As comparison, non-detection was found for the control stacking regions. 

\begin{table}[h]
\caption{X-ray count rates, X-ray luminosities and X-ray emissivities of stacked GCs.  }
\label{tbl-02}
\tiny
  \begin{center}\begin{tabular}{cccc}
  \hline\hline
Bands/Groups & net count rate & $L_{X}$ & $\varepsilon_{\rm X}$  \\
     &   ($10^{-6}~\rm cts s^{-1}{\rm GC}^{-1}$)   &  ($1.0\times10^{33}{\rm~erg~s^{-1}}~{\rm GC^{-1}} $)      &    ($1.0\times10^{27}{\rm~erg~s^{-1}}/M_{\odot}$)  \\
  \hline\noalign{\smallskip}
\multicolumn{4}{c}{All GCs}\\
F  & $7.4\pm2.2$& $5.3\pm1.6$ & $13.2\pm4.3$  \\
S  & $5.6\pm1.9$&$2.9\pm1.0$ &$7.3\pm2.5$ \\
H  & $1.7\pm1.1$&$1.9\pm1.3$ & $4.8\pm3.2$ \\
\hline
\multicolumn{4}{c}{Subgroups of GCs}\\
F,$\log~M>6.0$ & $11.9\pm6.4$& $8.4\pm4.5$ & $4.5\pm2.4$ \\
F,$\log~M<6.0$ & $4.4\pm2.3$& $3.1\pm1.6$ & $15.0\pm7.8$ \\
\hline
\multicolumn{4}{c}{MW GCs}\\
F,all GCs      & $-$ & $3.8\pm0.6$    & $7.3\pm2.7$ \\
\hline
F,$\log~M>6.0$ & $-$ & $14.4\pm1.0$ & $5.6\pm2.9$ \\
F,$\log~M<6.0$ & $-$ & $2.8\pm0.5$ & $14.6\pm9.3$ \\
\end{tabular}
\end{center}
Note.-Column 1: Stacking Bands or sub groups of GCs. Column 2:0.5-8 keV count rate. Column 3: 0.5-8 keV X-ray luminosity. Column 4: Specific X-ray emissivity of GCs.
\end{table}

In Figure 3, the count number profile of the merged C map were plotted (black dots). From the figure, the count distribution has a peak at the center pixels and drops to a near-constant value outside 3 pixels, which were considered as background (blue line). As a result, even an increment of source region to $10\times10$ pixel box does not result in any increase in net count rate or luminosity. Therefore, we conclude that our choice of the source region is reasonable. For comparison, the merged count profile of a randomly selected sample of 10 detected sources is also presented in Figure 2 (the profile was normalized so that it has the same count numbers at the center pixels with merged GCs). These 10 sources were randomly chosen within 8\arcmin~ of M\,31 center and have count rates within 10 times the detection limit. It is clear that both merged maps have similar count distribution: the count profiles drop to a quasi-constant level outside 3 pixels. We again conclude that the sizes of source and background region is reasonable.
\begin{figure}
	\centering
	\includegraphics[height=4in,width=5.4in]{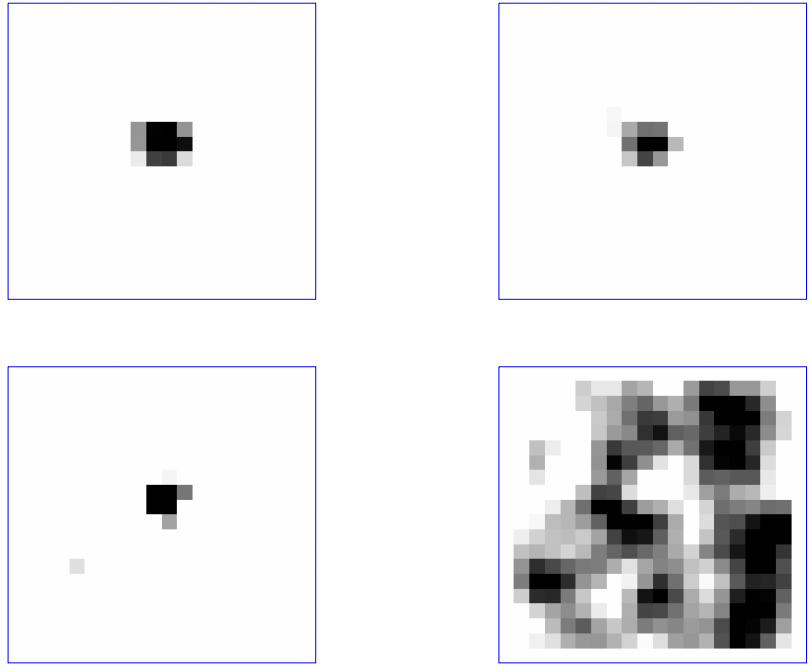}
\caption{Stacked count maps for sampled GCs in F band (upper left), S band (upper right), H band (lower left) and control field in F band (lower right). The maps are smoothed by a 2-pixel Gaussian kernel for enhanced illustration only.}
\label{Fig:03}
\end{figure}

The net count rates are then calculated as $C_{\rm N}/E$, where $E$ is the respective mean exposure time from the exposure map. The flux and luminosity for each band is calculated by assuming an absorbed power law spectrum. The slope $\Gamma$ of the spectrum was assumed to be $1.7$, which is typical for MW GCs and is consistent with the H/S count ratio of $0.4$. The count rate to flux conversion factor is then determined to be $1.25 \times 10^{-11} erg~{\rm s}^{-1} {\rm cts}~{s}^{-1}$ from the assumed power law spectrum. We choose the Poisson error to represent the uncertainties of the count numbers in source and background regions and calculate the other uncertainties accordingly.

As the next step, the specific X-ray emissivity $\varepsilon_{\rm X}$ of sampled GCs  is calculated by $\varepsilon_{\rm X}=L_{\rm X}/M_{\rm avg}=13.2\pm4.3\times10^{27}\rm erg~{\rm s}^{-1}~M^{-1}_{\odot}$, where $L_{\rm X}$ is the averaged X-ray luminosity of sampled GCs, and the averaged mass ($M_{\rm avg}$) of GCs were adopted from \citet{maj15} and the uncertainties of the GC masses have been included.  As the next step, the sampled GCs are divided into two groups according to their masses. Group 1 GCs have $\log~M/M_{\odot}>6.0$ and Group 2 GCs have $\log~M/M_{\odot}\leq6.0$. Stacking is then performed for each group and the results are also listed in Table \ref{tbl-02}. Apparently, more massive GCs on average have lower $\varepsilon_{\rm X}$ than less massive ones. But their X-ray luminosities are only marginally higher than those less massive ones. Additionally, stacking analysis are performed on sub groups of GCs according to their color (represented by $m_{\rm B}-m_{\rm V}$) or the distance to M\,31 center and no apparent dependences are found.

\section{Discussion \& Conclusion}
Our stacking has a valid detection on $\sim~5~\sigma$ level. In contrary, a similar stacking work by \citet{vul14} found no valid detection. Comparing with \citet{vul14}'s paper, we suspect that there are two possible reasons to be responsible. The first one is the astrometry correction step before merging more than 100 \textit{Chandra} observations. For example, we used \textit{reproject\_aspect} command in \textit{CIAO} to compare positions of detected sources and make corrections to the aspect files before merging \citep[We have noticed that the correction step was also included in a later paper by the same group of authors, e.g.,][]{vul16}. This correction can reduce up to 85\% source residuals (typically $\sim~0.2$ to $0.8$\arcsec) and could be crucial when merging more than 100 \textit{Chandra} observations. For comparison, such a step is missing in the description of \citet{vul14}, and the \textit{reproject\_obs} command were directly used to merge the observations before correcting astrometry in \citet{vul14}. Another possible reason could be the different star cluster catalog used in the two works. \citet{vul14} used PHAT year 1 cluster catalog \citep[see][]{joh12} for stacking, while the revised Bologna catalog is used in this work. The former catalog is concentrated on the disk region, while the latter catalog covers the whole M31 region. As a result, the two catalog contain GCs with different properties which could lead to different stacking results.

The average count rate of our sampled GCs reaches $7.4\pm2.2 \times 10^{-6} {\rm cts}~{\rm s}^{-1} {\rm GC}^{-1}$ in F band, which is an order of magnitude lower than the detection limit. The average $L_{\rm X}$ ($5.3 \pm 1.6\times10^{33}\rm erg~{\rm s}^{-1} {\rm GC}^{-1}$) of sampled GCs is consistent with the average $L_{\rm X}$ of the MW GCs ($ 3.8 \pm 0.6\times10^{33}\rm erg~{\rm s}^{-1} {\rm GC}^{-1}$, see \citealt{czq17}).

The (marginally) higher average X-ray luminosity of massive GCs than less massive ones is as was found in the MW GCs, and can be naturally explained by more primordial binary star systems in massive GCs\citep{czq17}.  However, their lower $\varepsilon_{\rm X}$ suggests that CVs and/or ABs are less abundant in massive GCs than less massive ones. A similar trend was also found in MW GCs \citep{czq17}. Based on a sample of 69 MW GCs, \citet{czq17} proposed that the binary-single and binary-binary stellar encounters are more efficient in massive GCs, which would destroy (the progenitors of) CVs and ABs more effectively and reduce their numbers which leads to lower $\varepsilon_{\rm X}$. Similar mechanism could also be responsible for the lower $\varepsilon_{\rm X}$ in M\,31 GCs. Unfortunately, we are unable to determine the stellar encounter parameter $\Gamma$ for M\,31 GCs due to the lack of knowledge on dynamical parameters of sampled GCs. We suggest further observations on M\,31 GCs to test this possibility.

Our measurements do not suggest a lower emissivity of M31 GCs than that of the MW field stars. The measured specific emissivity of M\,31 GCs ($\varepsilon_{\rm X}=13.2\pm4.3\times10^{27}\rm erg~{\rm s}^{-1}~M^{-1}_{\odot}$) is consistent with both that of MW GCs ($\varepsilon_{\rm X}=7.3\pm2.7\times10^{27}\rm erg~{\rm s}^{-1}~M^{-1}_{\odot}$, \citealt{czq17}), and of MW field stars ($\sim 12\times10^{27} {\rm~erg~s^{-1}~M_{\odot}^{-1}}$; \citealt{saz06,gec15}). Given the lower mean mass of M31 GCs comparing to MW GCs (see Fig 4), the apparent consistency may be broken by further investigation.

\begin{figure}
\centering
\includegraphics[height=4in,width=5in]{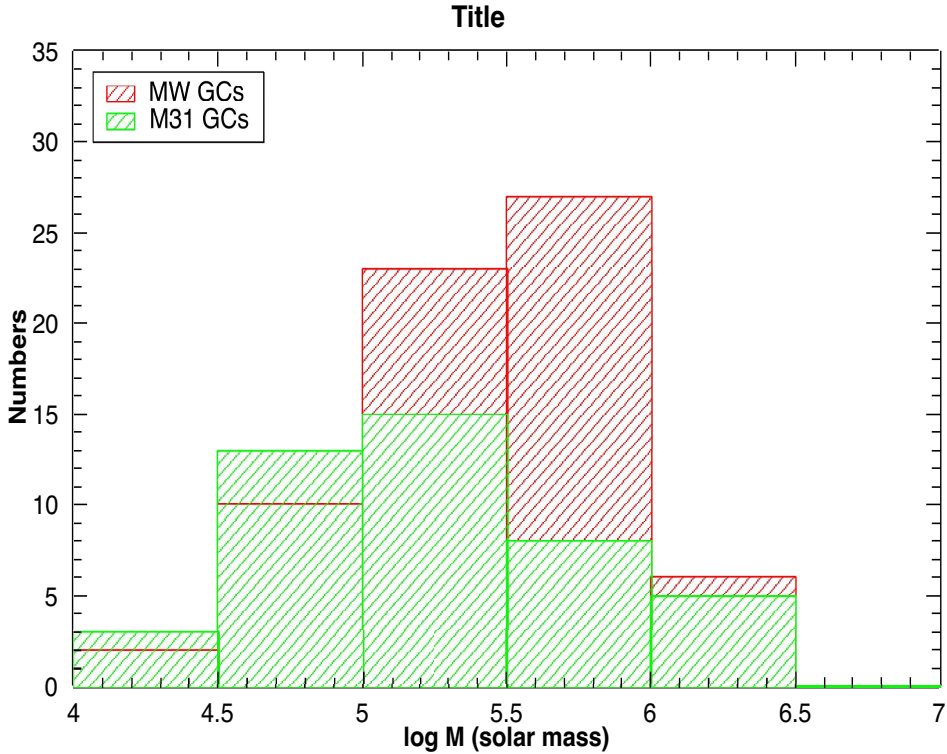}
   \caption{The mass distribution of sampled GCs. Green histograms are M\,31 GCs, and red ones are the MW GC sample from \citet{czq17}.}
   \label{Fig:04}
\end{figure}

Our results do not suggest the dependence of $L_{\rm X}$ per cluster on either the distance of GCs to M\,31 center, or $m_{\rm B}-m_{V}$, the color of GCs. It is possible that such dependences are relatively minor comparing to the mass and dynamical histories of M\,31 GCs, as was  found in MW GCs \citep{czq17}, which requires further study. What's more, the dynamical encounter rates and therefore the X-ray emitting source population in M\,31 GCs are not necessarily similar to those of the MW GCs, which will depends on future observations.  Last but not least, the sampled GCs in this research are located in M31 bulge, so they may have been experienced more dynamical evolution than typical globular clusters in M31\citep[e.g.,][]{aga13}, or be atypical of the overall population in other ways. Thus the overall emissivity of M31 GCs requires further investigations.

\acknowledgements
The authors thank the anonymous referee for constructive comments that helped improve this paper. This work is supported by National Science Foundation of China through grants NSFC-11303015, 11473010, and NSFC-11133004, the National Key Research and Development Program of China (2016YFA0400803), NSFC-11773015 and NSFC-11133001.

{\it Facilities:} \facility{CXO (ACIS)}.

\end{document}